\documentstyle[seceq,epsf]{ptptex}

\title{Analytical treatment of interacting Fermi gas in arbitrary 
dimensional harmonic trap
}

\author{
Hiroyuki {\sc Yoshimoto}\footnote{E-mail: hiroyuki@kh.phys.waseda.ac.jp} 
and Susumu {\sc Kurihara}
}

\inst{
Physics Department, Waseda University, 3-4-1 Okubo, Tokyo 169-8555, Japan}

\recdate{
}

\abst{
We study normal state properties of an interacting Fermi gas 
in an isotropic harmonic trap of arbitrary dimensions. 
We exactly calculate the first-order perturbation terms in the ground state 
energy and chemical potential, and obtain simple analytic expressions of the 
total energy and chemical potential. 
At zero temperature, we find that Thomas-Fermi approximation agrees well with 
exact results for any dimension even though system is dilute and small, i.e. 
when the Thomas-Fermi approximation is generally expected to fail. 
In the high temperature (classical) region, we find interaction energy decreases in proportion to $T^{-\frac{d}{2}}$,
 where $T$ is temperature and $d$ is dimension of the system. 
 Effect of interaction in the ground state in two and three-dimensional 
 systems is also discussed.  
}

\begin{document}

\maketitle

\section{Introduction}
Recently Fermi degeneracy in neutral atomic gases has been attained 
for $^{40}\rm K$ \cite{DeMarco} and $^6\rm Li$ \cite{Truscott,Schreck,O'Hara}.
In these systems, one can control physical parameters with relative ease.
Especially, the interaction strength can be varied in a wide range 
when Feshbach resonance is available \cite{Houbiers}. 
It is thus expected that the exciting phenomena of Fermion paring either in 
momentum space (Cooper pairing) or real space
 (BEC of preformed Fermion pairs),
leading to weak-coupling or strong-coupling superfluidity 
\cite{Stoof,Chiofalo,Holland,Ohashi}. 
This system is also characterized by the existence of trapping potential, 
which leads to discrete quantum eigenstates and shell structures. 
This makes analytical calculations rather difficult.
In many practical cases one uses Thomas-Fermi approximation for analytic
calculations. 
This method is justified in the case when system has sufficiently large number 
of particles. Several properties \cite{Search,Heiselberg,Bruun,Menotti}
such as single particle excitation and Cooper pair correlation is 
investigated in this approximation.
However it can't be always applicable when system is small 
and level discreteness play important roles.

The purpose of this paper is to investigate the normal ground state properties 
of dilute interacting Fermi gas in a trap with use of the exact eigenstate of 
the harmonic trap without resorting to the Thomas-Fermi approximation.
We give a systematic method to count up quantum number of harmonic potential 
when inter-particle interaction is contact type, and obtain a simple analytical expression of the total ground states energy and the chemical potential within 
the first-order perturbation theory.
Our method gives exact energy shift caused by interaction in arbitrary number 
of particle as for the lowest order term in interaction strength. Moreover, 
this method makes it possible to investigate effect of interaction to 
structure of degenerate oscillator shell exactly.
\section{GENERAL FORMALISM}
We consider a two-component Fermi gas with a short range interaction
in an isotropic harmonic potential. The two components (hyperfine state) are 
labeled by $\alpha=\uparrow\downarrow$ and assumed to have equal concentration.
We choose a unit system in which $\hbar=\omega=m=1$, where $m$ is the 
 atomic mass, and
 $\omega$ is the frequency of an isotropic trapping potential. 
 We start from the Hamiltonian $H=H_{0}+H_{\rm{int}}$, where
\begin{eqnarray} 
 H_{0}=\sum_\alpha\int d{\mathbf r}\,\psi_\alpha^\dagger({\mathbf r})
 (-\frac{1}{2}\frac{d^2}{d{\mathbf r}^2}+\frac{1}{2}{\mathbf r}^2)
 \psi_\alpha(\mathbf r) ,
\end{eqnarray}
is the one-particle part, $\psi_\alpha(\rm r)$ being the 
field operators of the component $\alpha$. 
\begin{eqnarray}
H_{\rm{int}}=g\int d\mathbf r\,\psi_\uparrow^\dagger(\mathbf r)
\psi_\downarrow^\dagger(\mathbf r)\psi_\downarrow(\mathbf r)
\psi_\uparrow(\mathbf r) ,
\end{eqnarray}
is the two-body interaction Hamiltonian, and $g$
is interaction strength. We don't use pseudopotential $\delta(\mathbf r)\partial_r r $ \cite{Castin,Martin,Bolda} , and use a delta function for simplicity. $H_{0}$ is easily diagonalized by writing the field operator as 
\begin{eqnarray}
\psi_\alpha({\mathbf r})=\prod_{i}^d\sum_{n^i}\frac{H_{n^i}(x_i)e^{
-\frac{1}{2}x^2}}{\sqrt{\pi^{\frac{1}{2}}n^i!2^{n^i}}}a_{\alpha n},
\end{eqnarray}
where $d$ is dimension of the system, $H_{n^i}(x_i)$ is the Hermite polynomial,
 and $n=(n^1,,)$ is quantum number,
for example $n=(n^1,n^2,n^3)$ stand for $d=3$, and $[a_{\alpha n},a_{\alpha' n'}^\dag]=\delta_{\alpha \alpha'}\delta_{n n'}$. Then $H_{\rm 0}$ and $H_{\rm{int}}$ is written 
in the form
\begin{eqnarray}
H_{0}=\sum_{i=1}^d \sum_{n^i \alpha}(n^i+\frac{1}{2})
a_{\alpha n}^\dagger a_{\alpha n}, 
\end{eqnarray}
\begin{eqnarray}
 H_{\rm int}=g\prod_{i=1}^d\sum_{n_{1}^i n_{2}^i n_{3}^i n_{4}^i}
 w(n_{1}^i,n_{2}^i,n_{3}^i,n_{4}^i)
 a_{\uparrow n_1}^\dagger a_{\downarrow n_2}^\dagger
 a_{\downarrow n_3}a_{\uparrow n_4}.
\end{eqnarray}
 Here, $w(n_1,n_2,n_3,n_4)$ is a function of quantum numbers of trapped system and is given by  
\begin{eqnarray}
w(n_1,n_2,n_3,n_4)=(-1)^{\frac{-n_1+n_2-n_3+n_4}{2}}\sqrt{\frac{(2n_1)!!(2n_3)!!}{(2n_2)!!(2n_4)!!}}\nonumber\\
\times\int_{0}^\infty\frac{dk}{\pi}k^{-n_1+n_2-n_3+n_4}
e^{-\frac{k^2}{2}}L_{n_1}^{n_2-n_1}(\frac{k^2}{2})L_{n_3}^{n_4-n_3}
(\frac{k^2}{2}) ,
\end{eqnarray}
where $ L_{n}^m(x)$ is the associated Laguerre polynomial. We can calculate this form using Fourier transformation \cite{Wonneberger,Xianlong}. Detailed properties of $w(n_1,n_2,n_3,n_4)$ is investigated in Ref.~\citen{Gleisberg}. 
\section{GROUND STATE PROPERTIES}
\subsection{\label{sec:level3}One-dimensional case}
We use the first-order perturbation theory, and obtain one-dimensional interaction energy for one-dimensional case in the form 
\begin{eqnarray}
  E_{\rm int}(n_F)=g\sum_{n=0,m=0}^{n_F}w(n,n,m,m).
\label{eq7}
\end{eqnarray}
 Here the function $ w(n,n,m,m)$ has complicated dependence on quantum numbers, showing considerable difference from momentum expansion ordinarily used to treat homogeneous system. We can nevertheless calculate Eq. (\ref{eq7}) exactly as shown in Appendix A, with the result
 \begin{eqnarray}
E_{\rm int}(n_F)=\frac{g}{\sqrt{2\pi}}\sum_{r=0}^{n_F}\Big(\frac{(2r+1)!!}{(2r)!!}\Big)^2\frac{(2n_F-2r-1)!!}{(2n_F-2r)!!}.
\label{eq8}
\end{eqnarray}
  Above summation can be reduced further to an integral, yielding a simple form of energy
\begin{eqnarray}
E(N)\cong \frac{N^2}{4}+\frac{4g}{3\pi^2}N^{\frac{3}{2}}.
\label{eq9}
\end{eqnarray}
Here, $ N=2N_\uparrow = 2N_\downarrow $ is total number of particles. Chemical potential is also calculated by the relation $\mu(N_\alpha)=(E(N_\alpha+1)-E(N_\alpha))/2$, yielding
\begin{eqnarray}
\mu(N_\alpha)\cong\frac{1}{2}+N_\alpha+\frac{2g}{\pi^2}\sqrt{2N_\alpha}.
\label{eq10}
\end{eqnarray}
Equations (\ref{eq9}), (\ref{eq10}) show that as $N$ increases, ratio $E_{\rm int}/E_{0}$ become small, thus one may see this system behaves like free particle system. However we should note the fact that even a weak interaction destroys Fermi-liquid nature in one dimension.    
\subsection{ Two-dimensional case}

In two-dimensional case, we have to deal with the degeneracy to calculate the energy. It turns out, however, as shown in Appendix B we can calculate the total energy exactly in this approximation when particle is filled up to Fermi level, with the result
\begin{eqnarray}
E(n_F)=\frac{1}{3}(n_F+1)(n_F+2)(2n_F+3) \nonumber \\
+\frac{g}{12\pi}(n_F+1)(n_F+2)(2n_F+3).
\label{eq11}
\end{eqnarray}
 The Fermi number $n_F$ and total number $N$ are related by $n_F=\frac{1}{2}(-3+\sqrt{4N+1})$ therefore, energy is given by
 \begin{eqnarray}
 E(N)=\frac{N}{3}\sqrt{4N+1}(1+\frac{g}{4\pi}).
  \end{eqnarray}
Both the single-particle term and the interaction term have the same dependence on number of particle, similar to the results for homogeneous system. Mean value of chemical potential $ \mu(N)=\frac{1}{2}\frac{\partial E}{\partial N}$ is given by
\begin{eqnarray}
\mu(N_\alpha)=\frac{12N_\alpha+1}{2\sqrt{8N_\alpha+1}}(1+\frac{g}{4\pi}).
 \label{eq13}
 \end{eqnarray}
In the following, we investigate the one particle energy shift. Let us suppose to add one more particle to system of which particles is filled up to Fermi level. Energy shift $ \xi(n_x,n_y)$  is generally a function of degenerate level $ n_x,n_y$  $(n_x+n_y=n_F+1)$. But in this case, it doesn't depend on the coordinate, in other words, interaction between particles in the different energy levels doesn't lifts degeneracy, and is given by 
\begin{eqnarray}
 \xi(n_x,n_y)=\frac{g}{4\pi}(n_F+1).
\label{eq14}
\end{eqnarray}
Note that interacting term in Eq. (\ref{eq13}) corresponds exactly to the above form. 
Next, we add one more particle having another hyperfine state to the system. Interaction between two particles on the same oscillator shell gives the deviation from Eq. (\ref{eq14}). This energy shift $\delta\xi(n_x,n_y;m_x,m_y)$ is given in Appendix C, with the result 
\begin{eqnarray}
 &&\delta\xi(n_x,n_y;m_x,m_y)\cong \nonumber \\
&&\left\{ \begin{array}{cccc}
 \frac{g}{2\pi\sqrt{\pi(n_F+1)}}(1+\frac{1}{\pi}\log 4(n_F+1) )& \mbox{$n_x=m_x=0$}\\
 \frac{g}{2\pi^2n_xn_y}(1+\frac{1}{\pi}\log 4n_x )(1+\frac{1}{\pi}\log 4n_y )& \mbox{$n_x=m_x \neq 0$}\\
 \frac{2g}{\pi^4(n_x-m_x)}K(i\sqrt{\frac{n_x}{n_x-m_x}})K(i\sqrt{\frac{m_y}{m_y-n_y}}) & \mbox{$n_x>m_x$}\\
 \frac{2g}{\pi^4(m_x-n_x)}K(i\sqrt{\frac{m_x}{m_x-n_x}})K(i\sqrt{\frac{n_y}{n_y-m_y}})  &\mbox{$n_x<m_x$   .}  
 \end{array}
 \right.
\label{eq15}
\end{eqnarray}
Here $m_x , m_y$ $(m_x+m_y=n_F+1)$ are coordinates of the second particle, and $K(x)$ is the complete elliptic integrals of the first kind. We find that the interaction between two particles in the same oscillator shell lifts degeneracy mentioned above. $ \mu(n_x,n_y;m_x,m_y) $ have large value when two coordinate is close to each other, especially it have maximum value when $ n_x=m_x=0 $ or $ n_y=m_y=0 $. Eq. (\ref{eq14}) and Eq. (\ref{eq15}) imply two particles in the same shell make bound pairs in the same quantum number, when interaction is attractive and system is sufficiently dilute. 
\subsection{Three-dimensional case}
Calculation can be done more or less similarly also in also three-dimensional 
case. The detail is given in Appendix D.
The result for total energy is given by
\begin{eqnarray}
E(n_F)=\frac{1}{4}(n_F+1)(n_F+2)^2(n_F+3) \nonumber \\
+\frac{g}{9(2\pi)^{\frac{3}{2}}}\sum_{r=0}^{n_F}
\Big(\frac{(2r+3)!!}{(2r)!!}\Big)^2\frac{(2n_F-2r+1)!!}{(2n_F-2r)!!},
\end{eqnarray}
\begin{eqnarray}
\cong\frac{3^{\frac{4}{3}}}{4}N^{\frac{4}{3}}+\frac{256\sqrt{6}g}{945\pi^3}
N^{\frac{3}{2}}.
\label{eq17}
\end{eqnarray}
Equation. (\ref{eq17}) can be also obtained from Thomas-Fermi approximation with weak interaction ($\mid g \mid\ll 1$) \cite{Heiselberg,Bruun,Menotti,Vichi} $E^{\rm TF}_{\rm int}(N)=\frac{g}{4}\int d^3x n(x)^2$, where $n(x)=\frac{2^{3/2}}{3\pi^2}[\mu-\frac{1}{2}x^2]^{3/2}$ is non-interacting particle density. It is possible to indicate the same procedure in one and two-dimensional case. As can be seen from Fig.3, exact first-order perturbation calculation for the interaction energies are very close to the Thomas-Fermi approximation. Although the Thomas-Fermi approximation is ordinarily correct in the system, which has sufficiently large number of particles, our results show this approximation is applicable to any dimensional trapped 
system, even if the system is dilute as this approximation is generally expected to fail.
Average chemical potential is given by
\begin{eqnarray}
\mu(N_\alpha)=(6N_\alpha)^\frac{1}{3}\Big(1+\frac{128\sqrt{2}g}{315\pi^3}(6N_\alpha)^\frac{1}{6}\Big).\end{eqnarray}
 We investigate energy shift more precisely as we have done on the two dimensional cases. One particle energy shift $\xi(n_x,n_y,n_z)$ is found to be 
  \begin{eqnarray}
   &&\xi(n_x,n_y,n_z)= \nonumber \\
   &&\frac{\sqrt{2}g}{4\pi^\frac{3}{2}}
   \frac{d^{n_F}}{d\beta^{n_F}}\Big(\beta^{\frac{1}{2}+n_F}F_{n_x}(\beta)F_{n_y}(\beta)F_{n_z}(\beta)\Big)\mid _{\beta =1}.
\label{eq19}
 \end{eqnarray}
 Where $n_x, n_y, n_z$ $(n_x+n_y+n_z=n_F+1)$ is coordinate of degenerate states and $F_{n}(\beta)$ stands for Hypergeometric function $ F(-n,\frac{1}{2},1;\frac{1}{\beta})$. 
 In this case, interaction lifts the degeneracy and Eq. (\ref{eq19}) has maximum value when two of the quantum numbers are zero. The maximum value is
 \begin{eqnarray}
  \xi(n_x,n_y,n_z)\cong \frac{\sqrt{2}g}{\pi^2}\Big(\frac{4}{9\pi}n_F^{\frac{3}{2}}+\frac{65}{192}n_F^{\frac{1}{2}}\Big).
 \label{eq20}
 \end{eqnarray}
 This form reproduces results of \cite{Heiselberg,Bruun}. Eq. (\ref{eq19}) have a minimum value at $ n_x\cong n_y\cong n_z$, namely one particle wave function assumes 
 spherically symmetric shape. 
We calculate the minimum value numerically and find it to be roughly 0.89 times lower than maximum value. 

\section{FINITE TEMPERATURE}
 
 At the finite temperature, number of particles and energy of $d$-dimensional systems is written in the form
 \begin{eqnarray}
N=2\sum_{n=0}^{\infty}D(n)f(n)
\label{eq21}
\end{eqnarray}
\begin{eqnarray}
 E= 2\sum_{n=0}^{\infty}D(n)(n+\frac{d}{2})f(n) \nonumber \\
 + g\prod_{i=1}^d\sum_{\scriptstyle n_i=0 \atop \scriptstyle m_i=0}^{\infty}
 h(n_i,m_i)f(n)f(m),
 \label{eq22}
 \end{eqnarray}
here, $f(n)$ is distribution function, $n=n_1+..n_d$, and $D(n)=\sum_{n_1+..n_d=n}1$ is density of state. We restrict our discussion to high temperature, i.e. well above the Fermi degeneracy. Then, we use Boltzmann distribution function and calculate Eq. (\ref{eq21}) and find to be
 \begin{eqnarray}
 N =2\frac{e^{\frac{1}{T}(\mu-\frac{d}{2})}}{(1-e^{-\frac{1}{T}})^d}\nonumber \\
 \cong2 T^d e^{\frac{1}{T}(\mu-\frac{d}{2})}.
 \label{eq23}
 \end{eqnarray}
 Here, we put on Boltzmann constant $k_B=1$ for simplicity. Using the Eq. (\ref{eq23}), we can calculate Eq. (\ref{eq22}) which is given in Appendix E with the result 
 \begin{eqnarray}
 E&=&\frac{d N}{2\tanh(\frac{2}{T})}+\frac{g}{4(2\pi)^{\frac{d}{2}}}N^2\tanh^{\frac{d}{2}}(\frac{2}{T}) \nonumber \\
 &\cong&dNT+\frac{gN^2}{4(4\pi T)^{\frac{d}{2}}}.
 \label{eq24}
 \end{eqnarray}
We have thus obtained both for $T=0$ and $T \to \infty$ limits. 
These results may be useful in constructing an interpolation formula
for finite temperatures. 
     
\section{SUMMARY}
 
 In summary, we have studied an interacting Fermi gas in arbitrary dimensional isotropic harmonic trap. We have used first-order perturbation theory. In contrast to the homogeneous system, first-order perturbation term has non-trivial contribution to the system. Dependence on the number of particles, dimension, and temperature is characteristic of the trapped system. We have calculated energy and chemical potential and found that Thomas-Fermi approximation is applicable when the system is close to the ground state, even if the system is dilute and small, i.e. when the Thomas-Fermi approximation is generally expected to fail, and we have found interaction energy decreases in proportion to $T^{-\frac{d}{2}}$, when the system is well above the Fermi degeneracy. 
\noindent

\section*{Acknowledgements}
This work was partly supported by COE Program "Molecular Nano-Engineering"
from the Ministry of Education, Science and Culture, Japan.

\appendix
\section{Derivation of Eq. (\ref{eq9}), Eq. (\ref{eq10})} 
 One easily find Eq. (\ref{eq7}) is written in the form
\begin{equation}
E_{\rm{int}}(n_F)=\frac{g}{\sqrt{2}\pi}\int^{\infty}_{0}dkk^{-\frac{1}{2}}e^{-k}\sum^{n_F}_{n=0}L_{n}(k)\sum^{n_F}_{m=0}L_{m}(k) 
\label{A1}
\end{equation}
We use the relation 
\begin{eqnarray}
&&L_{n}^a(x)=\sum_{r=0}^{n}\frac{\Gamma(a-b+r)}{r!\Gamma(a-b)}L_{n-r}^b(x),
\label{A2}
\end{eqnarray}
here, $\Gamma(r)$ is gamma function, we put on $a=1$ and $b=0$ and find to be
\begin{eqnarray}
E_{\rm{int}}(n_F)=\frac{g}{\sqrt{2}\pi}\int^{\infty}_{0}dkk^{-\frac{1}{2}}e^{-k}
L_{n_F}^1(k)^2. 
\label{A3}
\end{eqnarray}
We use Eq. (\ref{A2}) again, put on $a=1$ and $b=\-\frac{1}{2}$ and use the relation
\begin{eqnarray}
\int_{0}^\infty dxe^{-x}x^aL_{n}^a(x)L_{m}^a(x)=\frac{\Gamma(a+n+1)}{n!}\delta_{n,m}. 
\label{A4}
\end{eqnarray}
One easily finds Eq. (\ref{eq8}) from Eq.(\ref{A4}). We use Wallis's formula $(2r-1)!!/(2r)!!\cong 1/\sqrt{\pi r}$ and replace the summation with integral, then interaction term of Eq. (\ref{eq8}) is reduced to be
\begin{eqnarray}
\frac{4g}{\sqrt{2}\pi^2}\int^{n_F}_0dr\frac{r}{\sqrt{n_F-r}}. \label{A5}
\end{eqnarray}
We obtain Eq. (\ref{eq9}) directly from (A5). Interaction term of chemical potential is
written in the form
\begin{eqnarray}
\mu_{\rm int}(N_\alpha)&=&(E_{\rm int}(N_\alpha+1)-E_{\rm int}(N_\alpha))/2\nonumber \\
&=&g\sum_{n=0}^{n_F+1}f(n,n,n_F,n_F). \label{A6}
\end{eqnarray}
We use the relation Eq. (\ref{A2}) and Eq. (\ref{A4}) and find to be
\begin{eqnarray}
\mu_{\rm int}(N_\alpha)=\frac{g}{\sqrt{2\pi}}\sum_{r=0}^{n_F+1}\Big(\frac{(2r-1)!!}{(2r)!!}\Big)^2\nonumber \\
\times\frac{(2n_F+1-2r)!!}{(2n_F+2-2r)!!}(2r+1)\nonumber \\
\cong\frac{\sqrt{2}g}{\pi^2}\int^{n_F+1}_0dr\frac{1}{\sqrt{n_F+1-r}}.
\label{A7}
\end{eqnarray}
We obtain Eq. (\ref{eq10}) directly from (\ref{A7}).
\section{Derivation of two-dimensional energy}
Two-dimensional interaction energy is written in the form
\begin{eqnarray}
\!\!\!\!\!E_{\rm int}(N)=g\sum^{n_F}_{n=0}\sum^{n_F}_{m=0}\sum_{\scriptstyle n_x+n_y=n \atop \scriptstyle m_x+m_y=m}
h(n_x,m_x)h(n_y,m_y).
\label{B1}
\end{eqnarray}
Here, $h(n,m)$ represents $w(n,n,m,m)$. We use the relation
\begin{eqnarray}
\sum_{n_1+..+n_d=n}L_{n_1}(x_1)..L_{n_d}(x_d)=L^{d-1}_{n}(x_1+..+x_d).
\label{B2}
\end{eqnarray}
We put on $d=2$, then Eq. (\ref{B1}) is reduced to be
\begin{eqnarray}
E_{\rm int}(N)=g\sum^{n_F}_{n=0}\sum^{n_F}_{m=0}\int^{\infty}_{-\infty}dxdy
e^{-\frac{1}{2}(x^2+y^2)}\nonumber \\
\times L^1_{m}(\frac{x^2+y^2}{2})L^1_{n}(\frac{x^2+y^2}{2}).
\label{B3}
\end{eqnarray}
We replace above Cartesian coordinates valuable $(x,y)$ with polar coordinates valuables $(r,\theta)$ and use Eq. (\ref{A2}), then Eq. (\ref{B3}) is reduced to be
\begin{eqnarray}
E_{\rm int}(N)=\frac{g}{2\pi}\int^{\infty}_{0}d(\frac{r^2}{2})e^{-\frac{r^2}{2}}\Big(L_{n_F}^2(\frac{r^2}{2})\Big)^2.
\label{B4}
\end{eqnarray}
We apply Eq. (\ref{A2}) and Eq. (\ref{A4}) to above form, and we obtain Eq. (\ref{eq11}). 
Note that three-dimensional energy can be also calculated as well.
\section{ Derivation of Eq. (\ref{eq15}) }
Eq. (\ref{eq15}) is written in the form
\begin{eqnarray}
&&\delta\xi(n_x,n_y;m_x,m_y)= \nonumber \\
&&gh(n_x,m_y)h(n_F+1-n_x,n_F+1-m_y).
\label{C1}
\end{eqnarray}
We use Eq. (\ref{A2}) and Eq. (\ref{A4}) and find
\begin{eqnarray}
&&h(n,m)=\sqrt{2\pi}\sum^n_{r=0}\frac{(2r-1)!!}{(2r)!!} \nonumber \\
&&\times\frac{(2m-2n+2r-1)!!}{(2m-2n+2r)!!}\frac{(2n-2r-1)!!}{(2n-2r)!!}.
\label{C2}
\end{eqnarray}
If $ n=m $ above form is approximated to 
\begin{eqnarray}
h(n,n)&\cong&\sqrt{2\pi}\Big(\frac{1}{\sqrt{n\pi}}+\frac{1}{\pi^{\frac{3}{2}}}\int^n_1dr\frac{1}{r\sqrt{n-r}}\Big) \nonumber \\	
&\cong&\sqrt{\frac{2}{n}}\Big(1+\frac{1}{\pi}\log(4n)\Big),
\label{C3}
\end{eqnarray}
else if $m>n$ approximated to
\begin{eqnarray}
h(n,m)&\cong&\sqrt{2\pi}\frac{1}{\pi^{\frac{3}{2}}}\int^n_0dr\frac{1}{\sqrt{r(n-r)(m-n+r)}} \nonumber \\
&=&\frac{2\sqrt{2}}{\pi\sqrt{m-n}}K(i\sqrt{\frac{m}{m-n}}).
 \label{C4}
\end{eqnarray}
We use Eq. (\ref{C3}), (\ref{C4}) and obtain Eq. (\ref{eq15}).
\section{Derivation of Eq. (\ref{eq19}), (\ref{eq20}) }
We consider adding one more particle to three-dimensional system of which particles is filed up to Fermi level. Then, energy shift is 
\begin{eqnarray}
&&\xi(n_x,n_y,n_z)=\nonumber \\
&&\frac{g}{(2\pi)^3}\sum_{m=0}^{n_F}\sum_{m_x+m_y+m_z=m}
h(m_x,n_x)h(m_y,n_y)h(m_z,n_z)  \nonumber \\
&&=\frac{g}{\pi^3}\int^{\infty}_0drr^2e^{-\frac{1}{2}r^2}\int^{\frac{\pi}{2}}_0d\varphi
\int^{\frac{\pi}{2}}_0d\theta\sin\theta L^3_{n_F}(\frac{r^2}{2})\nonumber \\
&&\times L_{n_x}(\frac{r^2}{2}\sin^2\theta\cos^2\varphi)
L_{n_y}(\frac{r^2}{2}\sin^2\theta\sin^2\varphi)\nonumber \\
&&\times L_{n_z}(\frac{r^2}{2}\cos^2\theta).
\label{D1}
\end{eqnarray}
We expand the Laguerre polynomials and calculate angular integrals
\begin{eqnarray}
\int^{\frac{\pi}{2}}_0d\varphi
\int^{\frac{\pi}{2}}_0d\theta\sin\theta 
L_{n_x}(\frac{r^2}{2}\sin^2\theta\cos^2\varphi) \nonumber \\
\times L_{n_y}(\frac{r^2}{2}\sin^2\theta\sin^2\varphi)L_{n_z}(\frac{r^2}{2}
\cos^2\theta) \nonumber\\
=\frac{1}{4}\sum^{n_x}_{l_1=0}\sum^{n_y}_{l_2=0}\sum^{n_z}_{l_3=0}
(-1)^l \Big(\begin{array}{cc} n_x \\ l_x \end{array}\Big)
\Big(\begin{array}{cc} n_y \\ l_y \end{array}\Big)
\Big(\begin{array}{cc} n_z \\ l_z \end{array}\Big) \nonumber \\
\times \frac{r^l}{l_x!l_y!l_z!}
\frac{\Gamma(l_x+\frac{1}{2})\Gamma(l_y+\frac{1}{2})\Gamma(l_z+\frac{1}{2})}
{\Gamma(l+\frac{3}{2})},
\label{D2}
\end{eqnarray}
here, $ \big(\begin{array}{cc} n \\ l \end{array}\big) $ is binomial coefficient, and we puts on $ l=l_x+l_y+l_z $. Radial integral is also calculated and find to be 
\begin{eqnarray}
\int^{\infty}_0drr^{\frac{1}{2}+l}e^{-r}L^3_{n_F}(r)=\frac{\Gamma(\frac{5}{2}
-l+n_F)}{n_F!\Gamma(\frac{5}{2}-l)}\Gamma(\frac{3}{2}+l).
\label{D3}
\end{eqnarray}
From Eq. (\ref{D2}) and Eq. (\ref{D3}), we find 
\begin{eqnarray}
&&\xi(n_x,n_y,n_z)=\frac{g\sqrt{2}}{4\pi^3n_F!}
\sum^{n_1}_{l_1=0}b(n_x,l_x)\sum^{n_y}_{l_y=0}b(n_y,l_y)\nonumber \\
&&\times \sum^{n_z}_{l_z=0}b(n_z,l_z)\frac{\Gamma(\frac{5}{2}-l+n_F)}{\Gamma(\frac{5}{2}
-l)},
\label{D4}
\end{eqnarray}
here, $b(n,l)$ is $(-1)^l\big(\begin{array}{cc} n \\ l \end{array}\big)\
\Gamma(l+\frac{1}{2})/l! $. Eq. (\ref{eq19}) is derived from above form.
When $n_y=n_z=0$, Eq. (\ref{D4}) is reduced to be
\begin{eqnarray}
&&\xi(n_F+1,0,0)=(n_F+1)\frac{\sqrt{2}g}{2\pi ^{\frac{3}{2}}}\sum^{n_F+1}_{l=0}
\frac{(2l-1)!!}{(2l)!!} \nonumber \\
&&\times\frac{(2l-5)!!}{(2l)!!}\frac{(2n_F-2l+3)!!}{(2n_F+2-2l)!!} \label{D5}
\end{eqnarray}
We find that $\frac{(2l-1)!!}{(2l)!!}\frac{(2l-5)!!}{(2l)!!}$ decreases in proportion to $l^{-3}$, on the other hand  $\frac{(2n_F-2l+3)!!}{(2n_F+2-2l)!!}$ decreases in propotion to $(n_F-l)^{\frac{1}{2}}$ and have nearly constant value if $ l<<n_F$. Hence, we find
\begin{eqnarray}
&&\xi(n_F+1,0,0)\cong(n_F+1)\frac{\sqrt{2}g}{2\pi ^2}\sum^{n_F+1}_{l=0}
\frac{(2l-1)!!}{(2l)!!}\nonumber \\
&&\times\frac{(2l-5)!!}{(2l)!!}
(n_F^{\frac{1}{2}}+O(n_F^{-\frac{1}{2}})) \nonumber \\
&&\cong\frac{4\sqrt{2}}{9\pi^3}n_F^{\frac{3}{2}} + O(n_F^{\frac{1}{2}}).
\label{D6}
\end{eqnarray} 
Here, we use the relation $\pi\sum^{\infty}_{l=0}
\frac{(2l-1)!!}{(2l)!!}\frac{(2l-5)!!}{(2l)!!}=4/9$. We obtain the first term of Eq. (\ref{eq20}). Coefficient of second term is approximated by summation of Eq.(\ref{D5}) up to $l=2$. 
\section{Derivation of Eq. (\ref{eq24}) }

At the high temperature i.e. Boltzmann distribution is applicable, $d$-dimensional energy is given by
\begin{eqnarray}
E_{0}&=&2\sum^{\infty}_{n=0}D(n)(n+\frac{d}{2}) \label{E1} \\
&=&de^{\frac{1}{T}(\mu-\frac{d}{2})}\frac{(1+e^{-\frac{1}{T}})}
{(1-e^{-\frac{1}{T}})^{d+1}} \nonumber \\
&\cong& d e^{\frac{1}{T}(\mu-\frac{d}{2})}(1+e^{-\frac{1}{T}})T^{d+1}\label{E2}.\end{eqnarray}
Kinetic term of Eq. (\ref{eq24}) is obtained from Eq. (\ref{E2}) and Eq. (\ref{eq23}). On the other hand interaction term is given by
\begin{eqnarray}
&&E_{\rm int}=\frac{g}{(2\pi)^d}\sum^{\infty}_{n=0}\sum^{\infty}_{m=0}\sum_{\scriptstyle n_1+..n_d=n \atop \scriptstyle m_1+..m_d=m}e^{-\frac{1}{T}(n_1+..n_d-\mu +\frac{d}{2})}\nonumber \\
&&\times e^{-\frac{1}{T}(m_1+..m_d-\mu +\frac{d}{2})}\int_{-\infty}^{\infty}dx_1..dx_de^{-\frac{1}{2}(x_1^2+..x_d^2)}\nonumber \\
&&\times L_{m_1}(\frac{x_1^2}{2})..L_{m_d}(\frac{x_d^2}{2})L_{n_1}(\frac{x_1^2}{2})..L_{n_d}(\frac{x_d^2}{2}) \label{E3}.
\end{eqnarray}
We use the relation (\ref{B2}) then, above form is reduced to be
\begin{eqnarray}
&& E_{\rm int}=\frac{g}{(2\pi)^d}\frac{d\pi^{\frac{d}{2}}}{\Gamma(\frac{d}{2}+1)}\int^{\infty}_0 drr^{d-1}e^{-\frac{r^2}{2}}\nonumber \\
&& \times \Big (\sum^{\infty}_{n=0}e^{-\frac{n}{T}}L^{d-1}_n(\frac{r^2}{2})\Big)^2 . \label{E3}
\end{eqnarray}

We use Eq. (\ref{eq23}) and the relation 
\begin{eqnarray}
\sum^{\infty}_{n=0}t^{n}L^{d-1}_n(x)=\frac{e^{-\frac{xt}{1-t}}}{(1-t)^d},
\label{E4}
\end{eqnarray}
and find to be
\begin{eqnarray}
E_{\rm int}=\frac{g}{(2\pi)^d}\frac{d\pi^{\frac{d}{2}}}{\Gamma(\frac{d}{2}+1)}
N^2\int^{\infty}_0 drr^{d-1}e^{-\frac{r^2}{2\tanh(\frac{1}{2T})}}.
\label{E4}
\end{eqnarray}
We obtain Eq. (\ref{eq24}) from Eq. (\ref{E4}).

\begin{figure}
\centerline{\epsfbox{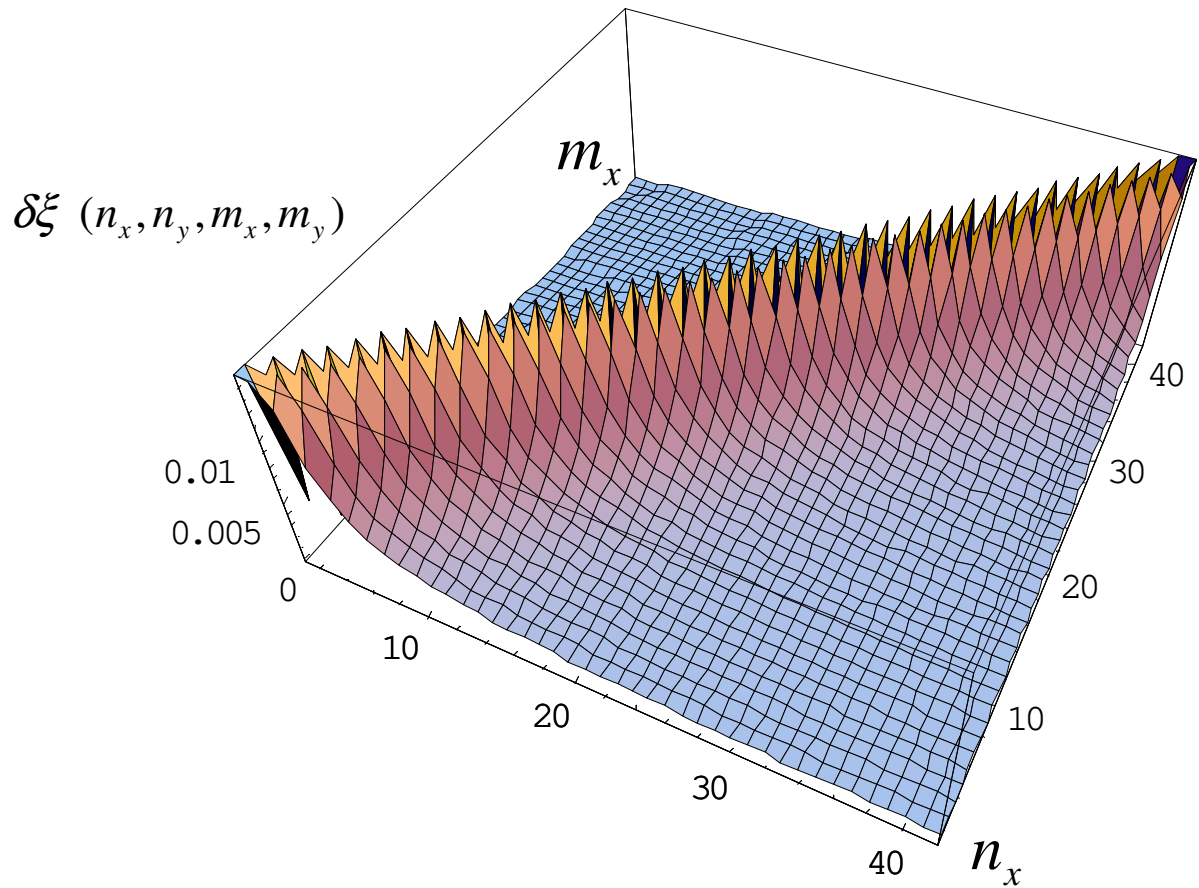}}
\caption{Energy shift caused by interaction between two particles  in the same energy level $n_F+1=41$ for $g=1$. $(n_x,n_y)=(n_x,n_F+1-n_x)$ and $(m_x,m_y)=(m_x,n_F+1-m_x)$ stand for two-dimensional degenerate states of the two particles.}
\end{figure}
 \begin{figure}[htpb]
\centerline{\epsfbox{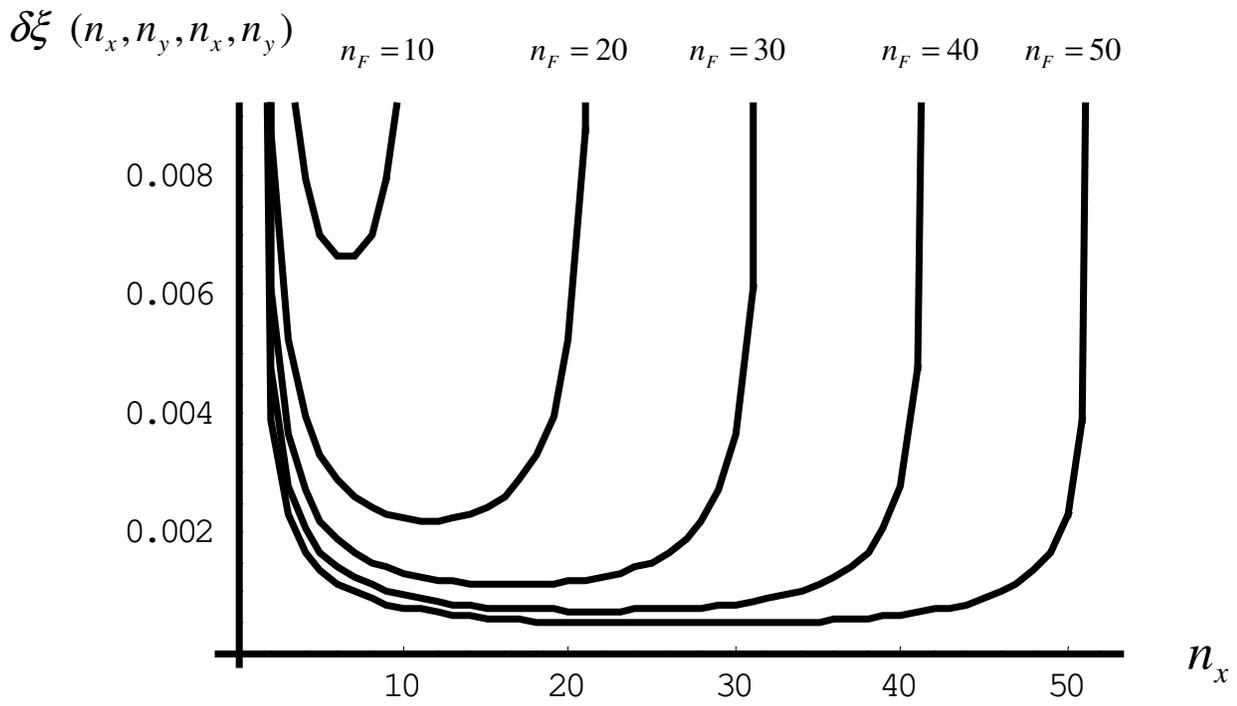}}
\caption{ Energy shift caused by interaction between two particles on the same two-dimensional degenerate states $(n_x,n_y)=(n_x,n_F+1-n_x)$ for g=1.
  }
\label{test5}
\end{figure}
\begin{figure}[htpb]
\centerline{\epsfbox{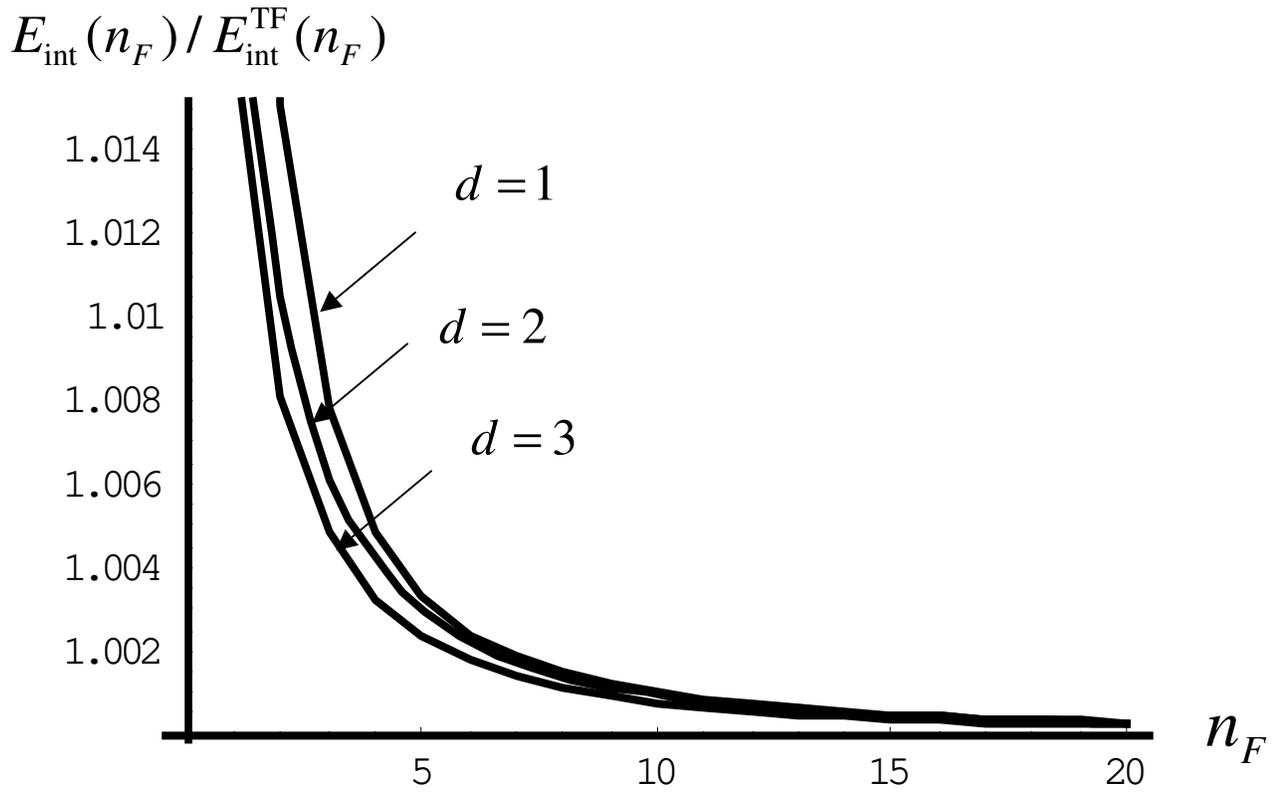}}
\caption{Ratio as a function of Fermi level $n_F$ for the interaction energy using Thomas-Fermi approximation $E_{\rm int}^{\rm TF}(n_F)$ and exact first-order perturbation. $d$ represents dimension of the system.}
\label{test3}
\end{figure}
\begin{figure}[htpb]
\centerline{\epsfbox{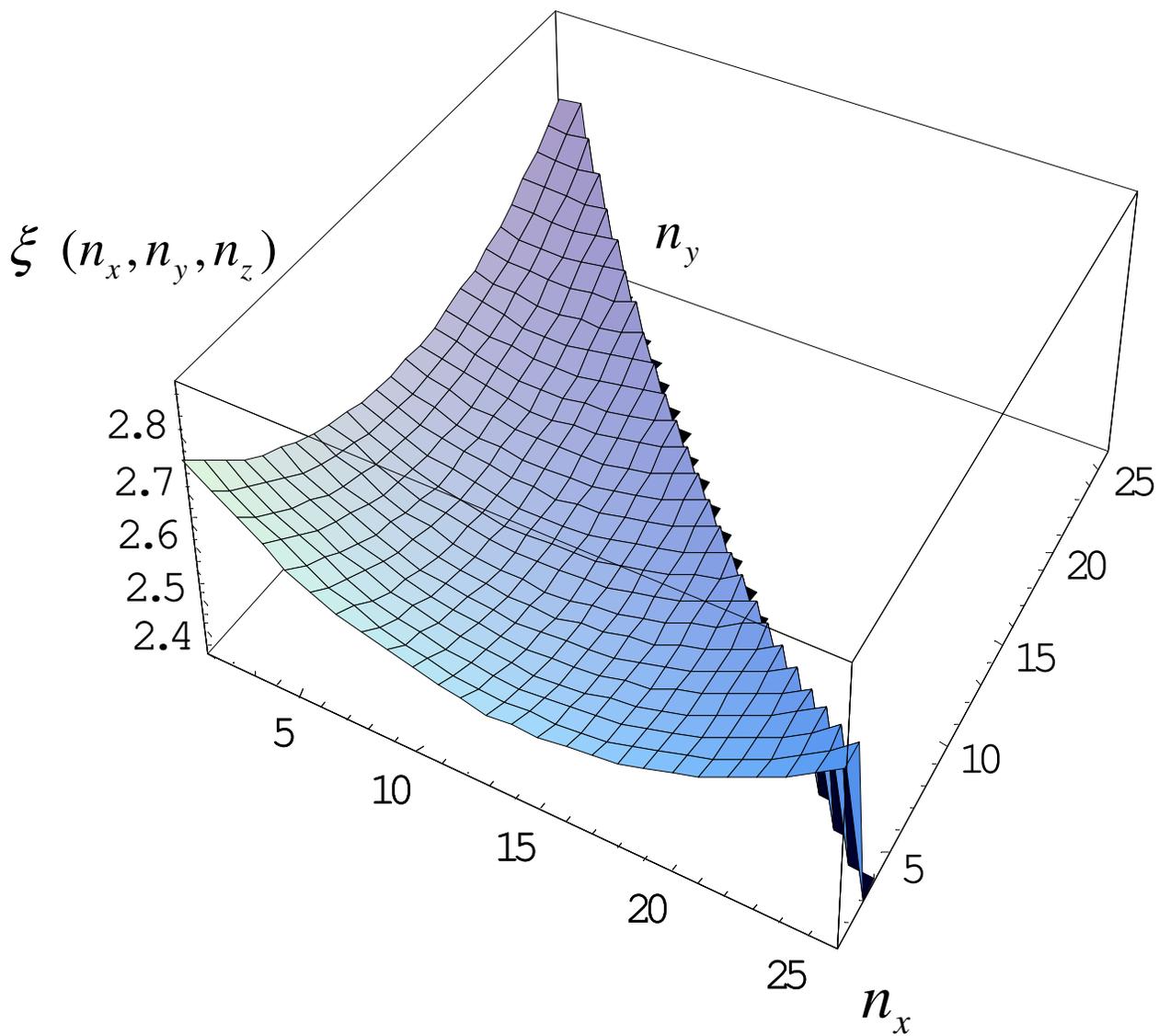}}
\caption{Single particle energy shift of the three-dimensional system caused by inter-shell interaction
for $ g=1$. $(n_x,n_y,n_z)=(n_x,n_y,n_F+1-n_x-n_y)$ is degenerate state of the particle. }
\end{figure}  
\end{document}